\documentclass[twocolumn,showpacs,preprintnumbers,superscriptaddress,amsmath,amssymb]{revtex4-2}
\usepackage{CJK}
\usepackage{amssymb}		
\usepackage{amsmath}
\usepackage{graphicx}
\usepackage{subfigure}
\usepackage[normalem]{ulem}
\usepackage{multirow}
\usepackage{appendix}
\usepackage{float}
\usepackage[usenames]{color}
\usepackage{bm}
\usepackage[colorlinks,linkcolor=blue,urlcolor=blue,citecolor=blue]{hyperref}
\usepackage{booktabs} 	     
\usepackage{leftidx}
\usepackage{soul,xcolor}
\setstcolor{red}
\newcommand\redsout{\bgroup\markoverwith{\textcolor{red}{\rule[0.5ex]{2pt}{0.4pt}}}\ULon}

\makeatletter

\newcommand{\Rmnum}[1]{\expandafter\@slowromancap\romannumeral #1@}
\makeatother

\usepackage{makecell}

\usepackage{tikz,xcolor,hyperref}

\definecolor{lime}{HTML}{A6CE39}
\DeclareRobustCommand{\orcidicon}{
	\begin{tikzpicture}
		\draw[lime, fill=lime] (0,0) 
		circle [radius=0.16] 
		node[white] {{\fontfamily{qag}\selectfont \tiny ID}};
		\draw[white, fill=white] (-0.0625,0.095) 
		circle [radius=0.007];
	\end{tikzpicture}
	\hspace{-2mm}
}
\foreach \x in {A, ..., Z}{%
	\expandafter\xdef\csname orcid\x\endcsname{\noexpand\href{https://orcid.org/\csname orcidauthor\x\endcsname}{\noexpand\orcidicon}}
}

\begin{document}
\begin{CJK*}{UTF8}{gbsn}	
	\title{Signatures of an $\alpha$ + core structure in $^{44}$Ti + $^{44}$Ti collisions at $\sqrt{s_{NN}}=5.02$ TeV by a multiphase transport model}
\author{Yu-Xuan Zhang(张宇轩)}
\affiliation{Key Laboratory of Nuclear Physics and Ion-beam Application (MOE), Institute of Modern Physics, Fudan University, Shanghai 200433, China}
\author{Song Zhang(张松)\orcidB{}}
\thanks{Email: song\_zhang@fudan.edu.cn}
\affiliation{Key Laboratory of Nuclear Physics and Ion-beam Application (MOE), Institute of Modern Physics, Fudan University, Shanghai 200433, China}
\affiliation{Shanghai Research Center for Theoretical Nuclear Physics， NSFC and Fudan University, Shanghai 200438, China}
\author{Yu-Gang Ma(马余刚)\orcidC{}}
\thanks{Email:  mayugang@fudan.edu.cn}
\affiliation{Key Laboratory of Nuclear Physics and Ion-beam Application (MOE), Institute of Modern Physics, Fudan University, Shanghai 200433, China}
\affiliation{Shanghai Research Center for Theoretical Nuclear Physics， NSFC and Fudan University, Shanghai 200438, China}

\begin{abstract}

It is important to understand whether $\alpha$-clustering structures can leave traces in ultra-relativistic heavy ion collisions. Using the modified AMPT model, we simulate three $\alpha$ + core configurations of $^{44}$Ti in $^{44}$Ti+$^{44}$Ti collisions at $\sqrt{s_{NN}}=5.02$ TeV as well as other systems with Woods-Saxon structures. One of these configurations has no additional constraint, but the other two have the Mott density edge $r_{\mathrm{Mott}}$ set as either a lower or upper bound on the cluster position $r_{\alpha}$ to check the influence of $\alpha$  dissolution.  This is the first time that the initial stage of the geometric properties in heavy-ion collisions has been configured using the traditional treatment of the nuclear structure.  We compare the radial nucleon density, multiplicity distribution, transverse momentum spectra, eccentricity, triangularity, elliptic flow and triangular flow of these six systems. $\alpha$ + core structures can alter all these observations especially in the most-central collisions, among which elliptic flow is the most hopeful as a probe of such structures.
\end{abstract}
\maketitle
\end{CJK*}	
	
\section{Introduction}\label{Introduction}
Predicted by quantum chromodynamics (QCD) \cite{andronic-decoding-2018}, the confinement of hadronic matter may be broken under extreme conditions at high temperature or density, resulting in a new matter state called quark-gluon plasma (QGP) \cite{SHURYAK1978150}. QGP contains deconfined quarks and gluons, providing us a chance to understand the essence of strong interaction on a new scale, which can help develop QCD further in return. Besides, the extreme conditions required by QGP are expected to exist in the early universe right after the Big Bang \cite{doi:10.1146/annurev.nucl.56.080805.140539}, so the investigation of QGP is also of great importance for revealing the universe origin. One of the important tools to gain such conditions is the ultra-relativistic heavy-ion collision, which is currently  performed by the CERN (European Organization for Nuclear Research) Large Hadron Collider (LHC) \cite{Lyndon-Evans-2008, acharya-multiplicity-2020, 201425} and the Relativistic Heavy Ion Collider (RHIC) at the Brookhaven National Laboratory \cite{HARRISON2003235, PhysRevLett.86.3500, PhysRevLett.88.022302}. Extensive studies are helpful in exploring the QCD phase structure and diagram \cite{Bzdak-PhysRep,LQCD-PhysRep,Qin-NST,Sun-NuclTech,Chen-NuclTech}. Among the results presented by different QGP probes, particle production \cite{PhysRevC.69.034909} and collectivity \cite{Ritter-2014} play instrumental roles in describing the evolution of collision systems. It is widely recognized that they reflect the initial state of system  and efforts to explain them usually involve modeling of the basic collision mechanism. 

Recently, the influences of nuclear structure are taken into account in this community by transport  or hydrodynamics models configured with initial nuclear structure~\cite{PhysRevLett.124.202301,PhysRevC.102.024901,PhysRevLett.131.022301,PhysRevC.105.044905,PhysRevLett.125.222301,PhysRevC.106.014906,PhysRevC.106.034909}. And these works suggest that there is  a potential window on investigating nuclear structure by relativistic heavy-ion collisions~\cite{Ma:2022dbh,bally2022imaging}.
Among various nuclear structures, the  $\alpha$-clustering structure inside nuclei is of particular interest.  The $\alpha$ cluster model, first proposed by Gamow \cite{Gamow}, has been demonstrated to be a powerful tool in describing nuclear structure \cite{BRINK1970143,Tohsaki2001,ZhouB}, $\alpha$ decay \cite{PhysRevC.99.034305,NST-1}, ground state bands \cite{PhysRevC.51.559} and so on. In this model, light nuclei could be thought to be made of $\alpha$ clusters as well as some nucleons or other smaller clusters \cite{BRINK1970143,PhysRevC.77.067301, PhysRevC.95.031303,Zhou}, while for heavy nuclei, only part of nucleons may cluster and a core could be formed by the remained nucleons \cite{PhysRevC.99.034305, PhysRevC.51.559,souza-alpha-2023}. 
The clustering effect is important to nuclear equations of state, nucleosynthesis and many other problems \cite{Qin,He:2014iqa,Ma-ML,Ma-NST}.  Various observables have therefore been proposed to study the clustering of nuclei in the heavy-ion reaction, such as collective flow \cite{PhysRevC.90.064902,PhysRevC.102.054907,Guo-PRC,Shi-NST},  multiplicity correlation \cite{Li:2022bpm,Li:2021znq} as well as giant resonance \cite{He:2014iqa,Huang1,Huang2,Huang3} etc. 
It is still an interesting question whether such cluster structures will form signatures in ultra-relativistic heavy-ion collisions. Some review on $\alpha$-clustering effects can be found in \cite{RMP,PhysRep,Ma:2022dbh,Ma-NuclTech}. For light nuclei, there is prediction \cite{PhysRevC.90.064902, PhysRevLett.112.112501,PhysRevC.95.064904} that clustering may lead to the variance of harmonic flow measures, implying a granular geometry preserved in ultra-relativistic heavy-ion collisions. Some positive simulation results \cite{PhysRevC.95.064904,PhysRevC.102.054907,PhysRevC.102.014910,Wang-PLB} have been reported under a multiphase transport (AMPT) model \cite{PhysRevC.72.064901}. However, for heavy nuclei, behaviors of $\alpha$-cluster structures remain unclear in ultra-relativistic collisions, and further investigation is required.

Among heavy nuclei, those with potential $\alpha$ + doubly magic core structures draw extra concentration since they match well with the binary $\alpha$ cluster model \cite{PhysRevC.51.559, SOUZA20198} and avoid a complex many-body problem \cite{PhysRevC.90.034304}. Through a modified AMPT model, we test this cluster model by simulating $^{44}$Ti + $^{44}$Ti collisions at $\sqrt{s_{NN}} = 5.02$ TeV with $\alpha$ + core ($\alpha$ + c) or Woods-Saxon (W-S) structures. We analyze the influence of this mode on yields and harmonic flow of major charged particles ($\pi^{\pm}$, $K^{\pm}$, $p$,  $\bar{p}$) and find possible signatures related to the $\alpha$ + $^{40}$Ca structure. Sec. \ref{METHODOLOGY} presents our approaches to acquiring the $\alpha$ + core simulation and analyzing the results. Results and discussion are shown in Sec. \ref{Result and discussion}, and the last section is the conclusion. 

\section{METHODOLOGY}\label{METHODOLOGY}
\begin{figure*}[htb]
	\includegraphics[width=17.8cm]{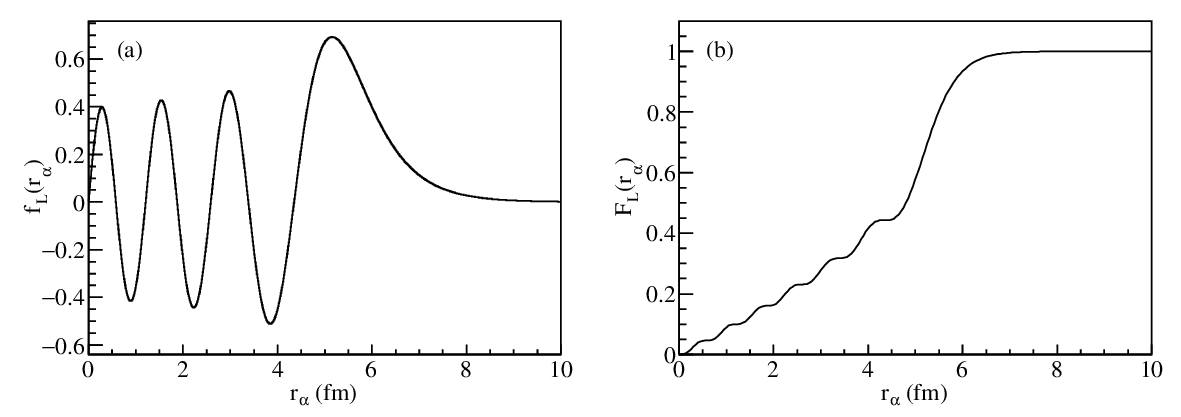}
	\caption{The normalized radial component $f_{L}(r_{\alpha})$ of the $\alpha$ cluster wave function for $^{44}$Ti at the ground state (a) and the cumulative  $r_{\alpha}$ distribution function $F_{L}(r_{\alpha})$ (b).} 
	\label{fig:flr}
\end{figure*} 

The AMPT model is a hybrid transport model aimed at simulating heavy ion collisions at RHIC and LHC energy \cite{PhysRevC.72.064901,Lin2021}. It contains several sub-models, such as the heavy-ion jet interaction generator (HIJING) model \cite{PhysRevD.44.3501}, Zhang's parton cascade (ZPC) model \cite{ZHANG1998276}, the Lund JETSET fragmentation model \cite{andersson-general-1983, SJOSTRAND199474}, a quark coalescence model and a relativistic transport (ART) model \cite{PhysRevC.52.2037}. In our work, heavy-ion collisions at the center of mass energy $\sqrt{s_{NN}}=5.02$ TeV will reach ultra-high temperature, leading to high QGP formation possibility. Hence, a string-melting version of the AMPT model is chosen for our simulation. In this model version, all excited strings are fragmented into partons and hadronized together with minijet partons, which is closer to the QGP case. 

In the AMPT model, the initial conditions of the collisions are provided by the HIJING model, where projectile and target nuclei are shaped into Woods-Saxon distributions and then the positions of constituent nucleons are set event by event. A 3-parameter Fermi function of the position $\mathbf{r}$ is used to describe Woods-Saxon distributions:
\begin{eqnarray}\label{Fermi}
	\begin{aligned}
		f(r)=&A\frac{1 +\omega r^2/c^{2}}{1+\exp[(r-c)/z]}.  
	\end{aligned}	
\end{eqnarray}
Here $A$ is the normalization factor, $c$ is the radius parameter, $z$ is the diffusion parameter, and $\omega$ is the newly added third parameter. If $\omega<0$, this function is cut off at $1 +\omega r^2/c^{2}=0$. For $^{40}$Ca, there are optimized parameters preset in the HIJING model, while for $^{44}$Ti, only automatically fitted data is available. $^{50}$Ti, a natural nuclei with a neutron magic number $N=28$, is considered as a singly closed shell core in the binary $\alpha$ cluster model \cite{souza-search-2017} and also not preset in the HIJING model. We additionally include it in our simulation to evaluate this fitting and test the system size effects. 

Instead of fixed cluster configurations common in light nucleus research \cite{PhysRevC.95.064904,PhysRevC.102.054907}, a non-localized $\alpha$ + $^{40}$Ca structure is introduced to modify the initial conditions by using the local potential model (LPM) \cite{PhysRevC.104.064301, PhysRevC.99.034305}. This structure features global motion of clusters and has well described the alpha condensate state of light nuclei like $^{12}$C \cite{PhysRevC.67.051306} and the $\alpha$ + core structure of heavy nuclei such as $^{20}$Ne \cite{PhysRevLett.110.262501} and $^{212}$Po \cite{PhysRevC.90.034304}. In LPM, we utilize the cluster position $r_{\alpha}$
 \cite{PhysRevC.104.064301, PhysRevC.99.034305}  to describe the $\alpha$ cluster potential $V(\mathbf{r_{\alpha}})$, which can be divided into the Coulomb potential $V_{C}(r_{\alpha})$, the centrifugal potential $V_{L}(r_{\alpha})$, and the nuclear potential $V_{N}(\mathbf{r_{\alpha}})$. With a uniformly charged spherical core assumption, $V_{C}(r_{\alpha})$ can be written as
\begin{eqnarray}\label{VCR}
	V_{C}(r_{\alpha})=\left\{
	\begin{aligned}
		&\frac{Z_{\alpha}Z_{c}e^{2}}{2R}\left(3-\frac{r_{\alpha}^2}{R^2}\right), &r_{\alpha}<R\\
		&\frac{Z_{\alpha}Z_{c}e^{2}}{r_{\alpha}}, &r_{\alpha}\geq R
	\end{aligned}
	\right.	
\end{eqnarray}
where $Z_{\alpha}$ and $Z_{c}$ are the charge numbers of the $\alpha$ cluster and the core, and $R$ is the core radius decided later by fitting $V_{N}(\boldsymbol{r_{\alpha}})$. The form of $V_{L}(r_{\alpha})$ is 
\begin{eqnarray}\label{VLR}
	V_{L}(r_{\alpha})=\frac{\hbar^2}{2\mu_{\alpha}r_{\alpha}^2}L(L+1),
\end{eqnarray}
with $L$ the azimuthal quantum number and $\mu_\alpha$ is the reduced mass in $\alpha$ two-body systems. Here we suppose that projectile and target nuclei are at their ground states, so $L$ is equal to zero and $V_{L}(r_{\alpha})$ can be neglected. As for $V_{N}(\boldsymbol{r_{\alpha}})$, we select a (1 + Gaussian) $\times$ (W.S. + W.S.$^{3}$) potential model \cite{SOUZA20198}, with which $V_{N}(\mathbf{r_{\alpha}})$ can be simply decided by the length $r_{\alpha}$. In this model there is 
\begin{eqnarray}\label{VNR}
	\begin{aligned}
	V_{N}(r_{\alpha})=&-V_{0} \left( 1 + \lambda e^{-r_{\alpha}^{2}/\sigma^2} \right) \left\{\frac{b}{1+e^{(r_{\alpha}-R)/a}}+\right.\\
	&\left.\frac{1-b}{[1+e^{(r_{\alpha}-R)/3a}]^3}\right\},
	\end{aligned}	
\end{eqnarray}
where $V_{0}$, $\lambda$, $a$ and $b$ are fixed parameters,  $R$ and $\sigma$ are free parameters for fitting. Here these parameters are the same as in Ref. \cite{SOUZA20198}, which well describes the ground state bands of $^{44}$Ti. Specifically, we set $V_{0} = 220~\mathrm{MeV}$, $a=0.65~\mathrm{fm}$, $b=0.3$, $\lambda=0.14$, $R=4.551~\mathrm{fm}$, and $\sigma=0.425~\mathrm{fm}$. Then with $\alpha$ decay energy $E_{\alpha}=-5.1271~\mathrm{MeV}$ \cite{Huang-2021}, we can gain the cluster's radial wave function $\varphi_{L}(\mathbf{r_{\alpha}})$ by numerically solve its stationary Schr$\ddot{\mathrm{o}}$dinger equation. Figure \ref{fig:flr} (a) shows the normalized radial component of the cluster wave function $f_{L}(r_{\alpha})=\varphi_{L}(\mathbf{r_{\alpha}})r_{\alpha}/Y_{LM}(\theta,\phi)$. The probability density of $r_{\alpha}$ can be gained from $\rho(r_{\alpha})=|f_{L}(r_{\alpha})|^2$ and the corresponding cumulative distribution function $F_{L}(r_{\alpha})$ is presented in Fig. \ref{fig:flr} (b).

With $F_{L}(r_{\alpha})$, we are able to insert an $\alpha$ cluster before core nucleons are placed. In the binary $\alpha$ cluster model, external influence on cluster or core nucleons is usually neglected, so here both $\alpha$ clusters and $^{40}$Ca cores have Woods-Saxon inner structures, and their parameters are simply decided by the HIJING preset data. For $\alpha$, it means $c=0.964~\mathrm{fm}$, $z=0.322~\mathrm{fm}$, and $\omega=0.517$ in Eq. (\ref{Fermi}). For $^{40}$Ca, there is $c=3.766~\mathrm{fm}$, $z=0.586~\mathrm{fm}$, and $\omega=-0.161$. According to Ref. \cite{PhysRevC.90.034304}, the Pauli blocking may lead to the dissolution of $\alpha$ clusters at the Mott density $\rho_{\mathrm{Mott}}=0.02917~\mathrm{fm^{-3}}$, which is $r_{\mathrm{Mott}}=4.498~\mathrm{fm}$ for $^{40}$Ca. This Mott density of a nucleus is given by the maximum density at which a nucleus of zero momentum can still be bound \cite{Ropke,WangRui}. That is to say, $\alpha$ clusters may not always exist in the $\alpha$ + core nuclei even if the model fully matches the reality. As pointed out in Refs. \cite{PhysRevC.99.034305,PhysRevC.101.024316,PhysRevC.93.011306}, the $\alpha$ cluster preformation probability $P_{\alpha}$ can be estimated by calculating its distribution outside $r_{\mathrm{Mott}}$. And according to these references, $P_{\alpha}$ usually has a magnitude of a few tenths, which can be high enough to form observable alteration. Shown in Fig. \ref{fig:flr} (b), about 55\% of clusters are in $r_{\alpha}>r_{\mathrm{Mott}}$ for the classical binary cluster model, which is acceptable considering the existing results. Since the clustering state is not completely dominant, we also test the influence of cluster dissolution by cutting off $r_{\alpha}$ at $r_{\mathrm{Mott}}$ like in $P_{\alpha}$ estimation in addition to the widely adopted no dissolution case. Though $r_{\alpha}<r_{\mathrm{Mott}}$ case results in the breaking of the nuclear matter saturation as discussed following, the simulation can be a comparison for $\alpha$ cluster effect with other cases.

\begin{figure}[!t]
	\includegraphics[width=8.6cm]{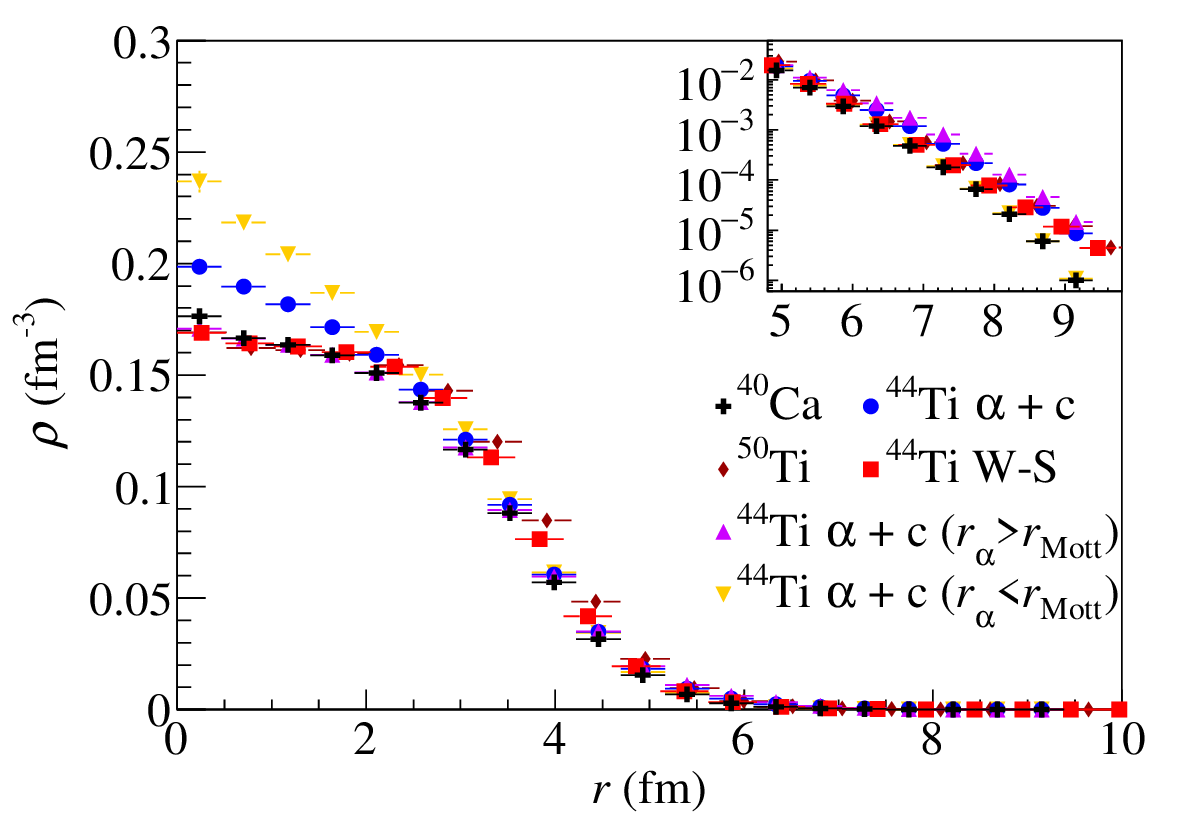}
	\caption{The nucleon density $\rho$ of projectile and target nuclei shown linearly in the main graph and logarithmically in the top right as a function of the distance $r$.} 
	\label{fig:r}
\end{figure}

\begin{figure}[tb]
	\includegraphics[width=8.6cm]{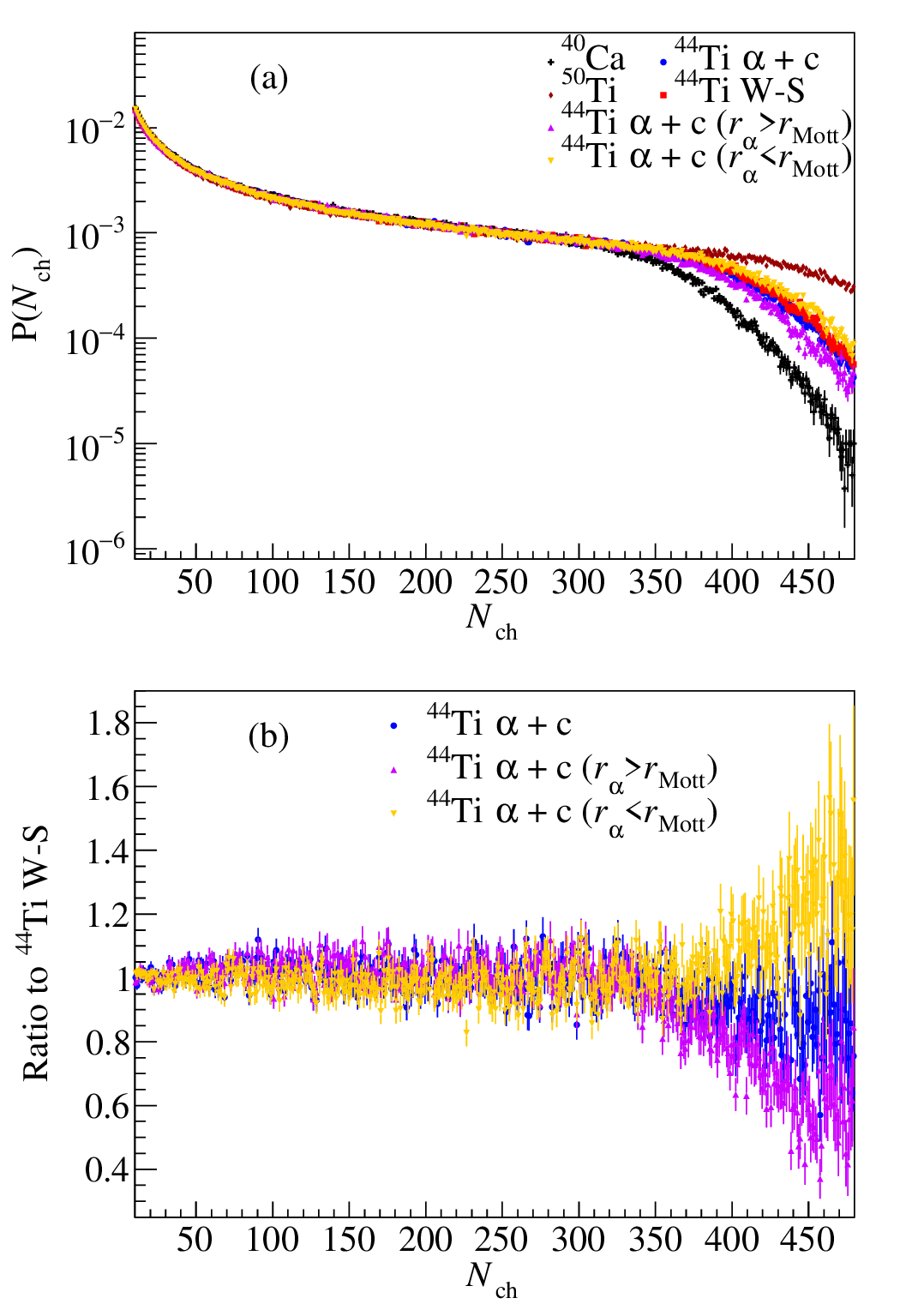}
	\caption{The multiplicity distribution of major charged particles within $|\eta|<0.5$ (a) and ratios to the Woods-Saxon structure in $^{44}$Ti + $^{44}$Ti systems (b).} 
	\label{fig:multi_hist}
\end{figure}

To characterize the initial geometry, we utilize the eccentricity $\varepsilon_{n}$ of participant partons and the anisotropic flow $v_{n}$ of final charged particles. $\varepsilon_{n}$ is a direct description of spatial anisotropy. In   specific heavy-ion collision, it can be defined as \cite{PhysRevC.102.014910}
\begin{eqnarray}\label{eccentricity}
	\varepsilon_{n}=\frac{\sqrt{\left\langle r^{n}\cos(n\varphi_{part}) \right\rangle^2+\left\langle r^{n}\sin(n\varphi_{part}) \right\rangle^2}}{\left\langle r^{n} \right\rangle},
\end{eqnarray}
where $\varphi_{part}$ is the azimuthal angle and  $r$ the position. $\varepsilon_{2}$ is usually called eccentricity and $\varepsilon_{3}$ is called triangularity. $\left<\cdots\right>$ denotes average over participant nucleons here, but later $\left<\varepsilon_n\right>$ means average over events. Hydrodynamics demonstrates a picture that the initial asymmetry in coordinate space will transfer to the final momentum space~\cite{NSTSong2017}. And we should point out the initial geometry asymmetry in the collisions includes the intrinsic structures and the overlapped region of the colliding nuclei, as well as fluctuation. In actual collision experiments, the participant parton positions are undetectable due to quark confinement, so description from collective flow reflecting anisotropy of particles at the final state is also necessary. On the other hand, the investigation of the final momentum space will disclose the properties of the initial stage and in further the intrinsic structure of nuclei. 

The anisotropic components of collective flow can be characterized by the Fourier expansion of particle momentum distribution \cite{PhysRevC.58.1671,Ma-FDU,ZhuLL,SSS,WangH,NST-ZhangB}:
\begin{eqnarray}\label{Fourier momentum distribution}
	E\frac{\mathrm{d}^{3}N}{\mathrm{d}^{3}p}=\frac{1}{2\pi}\frac{\mathrm{d}^{2}N}{p_{\mathrm{T}}\mathrm{d}p_{\mathrm{T}}dy}\left\{ 1+\sum_{n=1}^{\infty}2v_{n}\cos[n(\varphi-\Psi_{n})]\right\}.
\end{eqnarray}
Here $E$ is the particle energy, $p_{\mathrm{T}}$ is the transverse momentum, and $y$ is the rapidity. The Fourier expansion coefficient $v_{n}$ is the $n$-th order of the anisotropic flow and $\Psi_{n}$ is the corresponding event plane angle. Among all orders of flow, the elliptic flow $v_{2}$ and the triangular flow $v_{3}$ draw extra attention since they represent the initial collision geometry and its fluctuations, respectively \cite{2017193}. A common way to extract anisotropic flow is the cumulant method \cite{PhysRevC.83.044913}, which allows us to build multi-particle azimuthal correlations without looping over all particle multiplets. In a two sub-event case, with $\boldsymbol{Q}_n=\sum_{i=1}^{M} e^{in\varphi_{i}}$ storing the azimuthal angle information of $M$ particles, the two-particle correlation and its average over all events in this method can be written as \cite{PhysRevC.96.034906}
\begin{eqnarray}\label{2-correlation}
	\begin{aligned}
		\left\langle 2\right\rangle_{a|b}&=\frac{\boldsymbol{Q}_{n,a}\boldsymbol{Q}^{*}_{n,b}}{M_{a}M_{b}},\\
		\left\langle\left\langle 2\right\rangle\right\rangle_{a|b}&=\frac{\Sigma_{events} M_{a}M_{b}\left\langle 2\right\rangle_{a|b}}{\Sigma_{events} M_{a}M_{b}},
	\end{aligned}	
\end{eqnarray}
where $a$ and $b$ are symbols for these two sub-events. The corresponding $n$-th order flow can be written as
\begin{eqnarray}\label{2-cumulant-flow}
	v_{n}^{a|b}\{2\}=\sqrt{c_{n}^{a|b}\{2\}}=\sqrt{\left\langle\left\langle 2\right\rangle\right\rangle_{a|b}},	
\end{eqnarray}
where $c_{n}^{a|b}\{2\}$ is the two-particle cumulant. Here we assume the flow in two sub-events is the same, which is reasonable in a symmetric system with the same sub-event kinetic windows.

We can also generate two-particle correlation directly to extract collective flow \cite{Ma-PRC,2017193,2013213}. For trigger-associated particle correlation, we can go over all particle pairs and gain a per-trigger-particle associated yield, 
\begin{eqnarray}\label{direct-correlation}
	\begin{aligned}
		\frac{1}{N_{trig}}\frac{\mathrm{d}^{2}N^{pair}}{\mathrm{d}\Delta\eta \mathrm{d}\Delta\varphi } &= B(0,0)\frac{S(\Delta\varphi,\Delta\eta)}{B(\Delta\varphi,\Delta\eta)} ,\\
		S(\Delta\varphi,\Delta\eta) &= \frac{1}{N_{trig}}\frac{\mathrm{d}^{2}N^{same}}{\mathrm{d}\Delta\varphi \mathrm{d}\Delta\eta},\\
		B(\Delta\varphi,\Delta\eta) &= \frac{1}{N_{trig}}\frac{\mathrm{d}^{2}N^{mix}}{\mathrm{d}\Delta\varphi \mathrm{d}\Delta\eta},
	\end{aligned}	
\end{eqnarray}
where $\Delta\varphi$ and $\Delta\eta$ are the differences in $\varphi$ and pseudorapidity $\eta$ of the pair, and $N_{trig}$ is the trigger particle yield. $S(\Delta\phi,\Delta\eta)$ is the per-trigger-particle pair yield in the same event. $B(\Delta\phi,\Delta\eta)$ is generated by pairing trigger particles with associated particles from other events. The Fourier expansion of associated yields is 
\begin{eqnarray}\label{Fourier-direct-correlation}	
		\frac{1}{N_{trig}}\frac{\mathrm{d}N^{pair}}{\mathrm{d}\Delta\varphi } = \frac{N_{asso}}{2\pi}\left[1+\sum_{n=1}^{\infty}2V_{n\Delta}\cos(n\Delta\varphi)  \right],
\end{eqnarray}
where $V_{n\Delta}$ is the coefficient and $N_{asso}$ is the associated particle yield. With non-flow correlation reduced, there is 
\begin{eqnarray}\label{2-direct-correlation-Flow}
	V_{n\Delta}=v_{n}^{asso}v_{n}^{trig}, 
\end{eqnarray}
and
\begin{eqnarray}\label{2-direct-correlation-flow}
	v_{n}^{corr}=\sqrt{V_{n\Delta}}~~(v_{n}^{asso}=v_{n}^{trig}),
\end{eqnarray}
where $v_{n}^{asso}$, $v_{n}^{trig}$ and $v_{n}^{corr}$ are the flow of associated particles, triggers, and the entire correlation system respectively. In practice a $\Delta\eta$ gap is applied to Eq.(\ref{direct-correlation}) while projected into one-dimensional correlation function of Eq.(\ref{Fourier-direct-correlation}), which is considered to reduce non-flow effectively.

\begin{figure}[tb]
	\includegraphics[width=8.6cm]{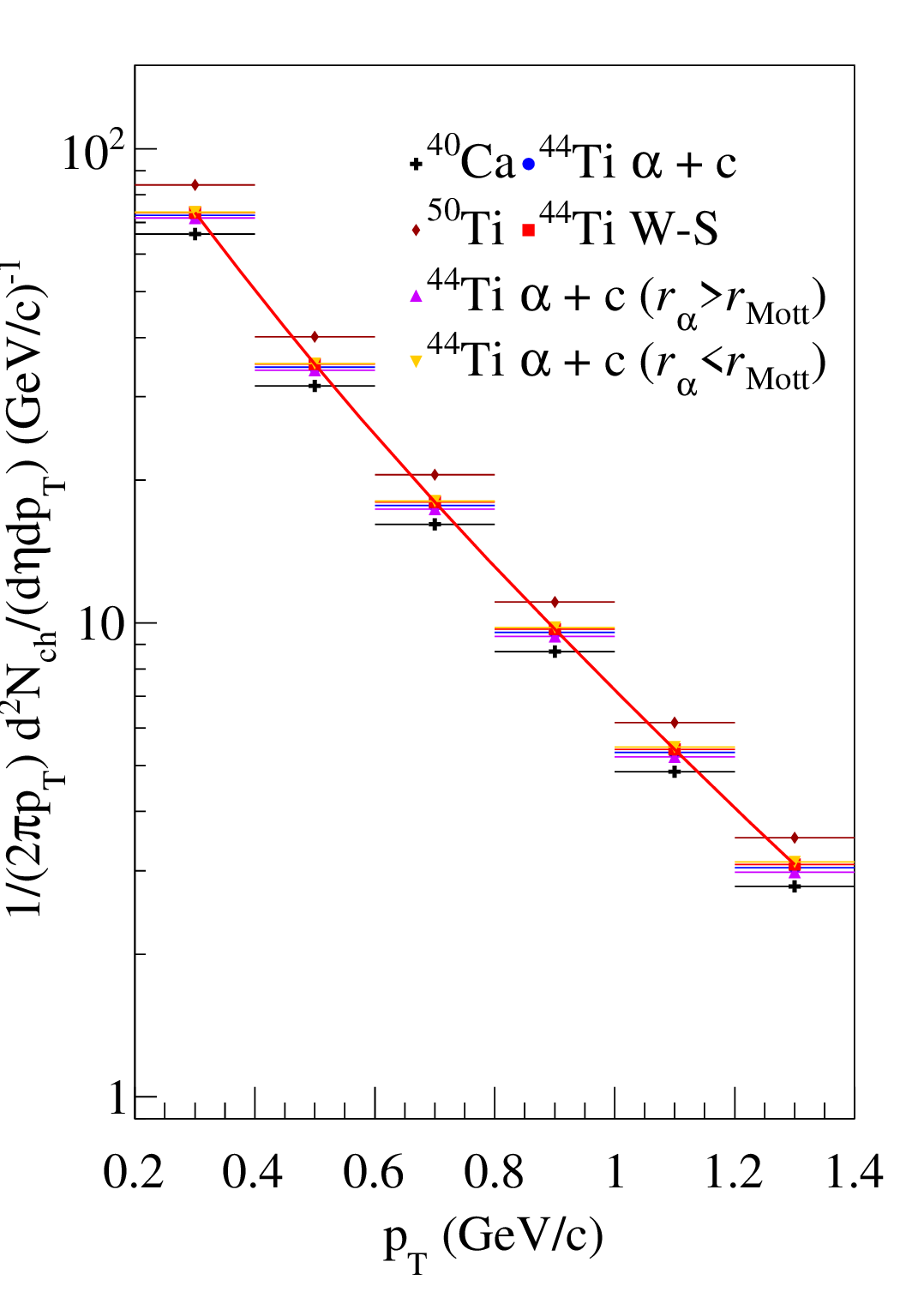}
	\caption{The transverse momentum spectra of major charged particles within $|\eta|<0.5$ for integrating all centralities. The red line connected with red squares is plotted for the case of W-S as a baseline for comparing other $^{44}$Ti + $^{44}$Ti systems. The same situation is presented in following figures.} 
	\label{fig:sp}
\end{figure} 

\section{Result and discussion}\label{Result and discussion}

In this work, three collision systems of Woods-Saxon nuclei and three of $\alpha$ + core $^{44}$Ti are simulated by the string-melting AMPT model version 2.26t7b. The first three systems are $^{50}$Ti + $^{50}$Ti, $^{40}$Ca + $^{40}$Ca, and $^{44}$Ti + $^{44}$Ti. In other three systems, one has no $r_{\alpha}$ constraint, one follows $r_{\alpha}>r_{\mathrm{Mott}}$, and the last one follows $r_{\alpha}<r_{\mathrm{Mott}}$. Each system has $8 \times 10^{5}$ events, which are uniformly divided into ten centrality classes according to the multiplicity of major charged particles within $p_{\mathrm{T}}>0.2~\mathrm{GeV/c}$ and $|\eta|<0.5$ ($N_{\mathrm{ch}}$).

To assess the initialization configuration, we first extract the radial nucleon density of projectile and target nuclei, which is presented in Fig. \ref{fig:r}. The central nucleon density is almost the same for Woods-Saxon $^{40}$Ca,  $^{50}$Ti and $^{44}$Ti, showing typical nuclear matter saturation. However, this feature is broken for nuclei with a classical $\alpha$ + core structure. Further density increase appears for $r_{\alpha}<r_{\mathrm{Mott}}$, while for $r_{\alpha}>r_{\mathrm{Mott}}$, the saturation is restored, indicating that this breaking is due to clusters neglecting Pauli blocking. This negligence widely exists in energy level calculation, where the result is mainly drawn from stationary Schr$\ddot{\mathrm{o}}$dinger equations regardless of the cluster dissolution \cite{SOUZA20198,PhysRevC.104.064301}. Some progress has been made in Ref. \cite{PhysRevLett.110.262501}, where cluster motion avoids mutual overlap due to the Pauli blocking effect. However, this just falls in the $r_{\alpha}>r_{\mathrm{Mott}}$ case and the cluster dissolution is not directly handled. So future optimization to introduce such transition is still necessary in relative areas. Another difference is in the peripheral area. For Woods-Saxon nuclei, the nucleon density decreases exponentially in peripheral areas, but for $\alpha$ + core structures, the clusters introduce another density enhancement. This time the Mott density constraint concentrates more clusters outside and hence leads to higher peripheral density than the classical cluster model's. 

Seeing such density distribution differences, we expect alteration related to nuclear structures in particle production. Fig. \ref{fig:multi_hist} (a) reports the probability distribution of $N_{\mathrm{ch}}$ as $\mathrm{P}(N_{\mathrm{ch}})$. Distributions of all systems look alike at low $N_{\mathrm{ch}}$, reflecting the low possibility of clusters appearing in collision zones and similar surface behaviors of Woods-Saxon nuclei or cores. But when $N_{\mathrm{ch}}$ is high, the influence of nuclear structures starts to play an important role in the multiplicity distribution. A $\mathrm{P}(N_{\mathrm{ch}})$ ratio comparison is presented for $^{44}$Ti + $^{44}$Ti systems in Fig. \ref{fig:multi_hist} (b), where clear depression is seen in $r_{\alpha}>r_{\mathrm{Mott}}$ at high $N_{\mathrm{ch}}$, while enhancement is shown instead when $r_{\alpha}<r_{\mathrm{Mott}}$. The mixed result of free $\alpha$ tends to decrease slightly since there are a bit more clusters outside the cores as discussed in Sec. \ref{METHODOLOGY}. We also present the transverse momentum spectra of major charged particles within $|\eta|<0.5$ for integrating  all centralities in Fig. \ref{fig:sp}, in which the red line connected with red squares is plotted for the case of W-S as a baseline (the same in the following figures).  The distinction due to nuclear structures is rather small but still exists this time, which is consistent with Fig. \ref{fig:multi_hist} (b) where the deviation only appears in the ultra-high $N_{\mathrm{ch}}$ events and is lower than one order of magnitude. This behaviour is similar to the multiplicity distribution and ratio in isobar collisions~\cite{PhysRevC.105.014901,PhysRevC.106.014906} while considering the neutron skin effect.

\begin{figure}[tb]
	\includegraphics[width=8.6cm]{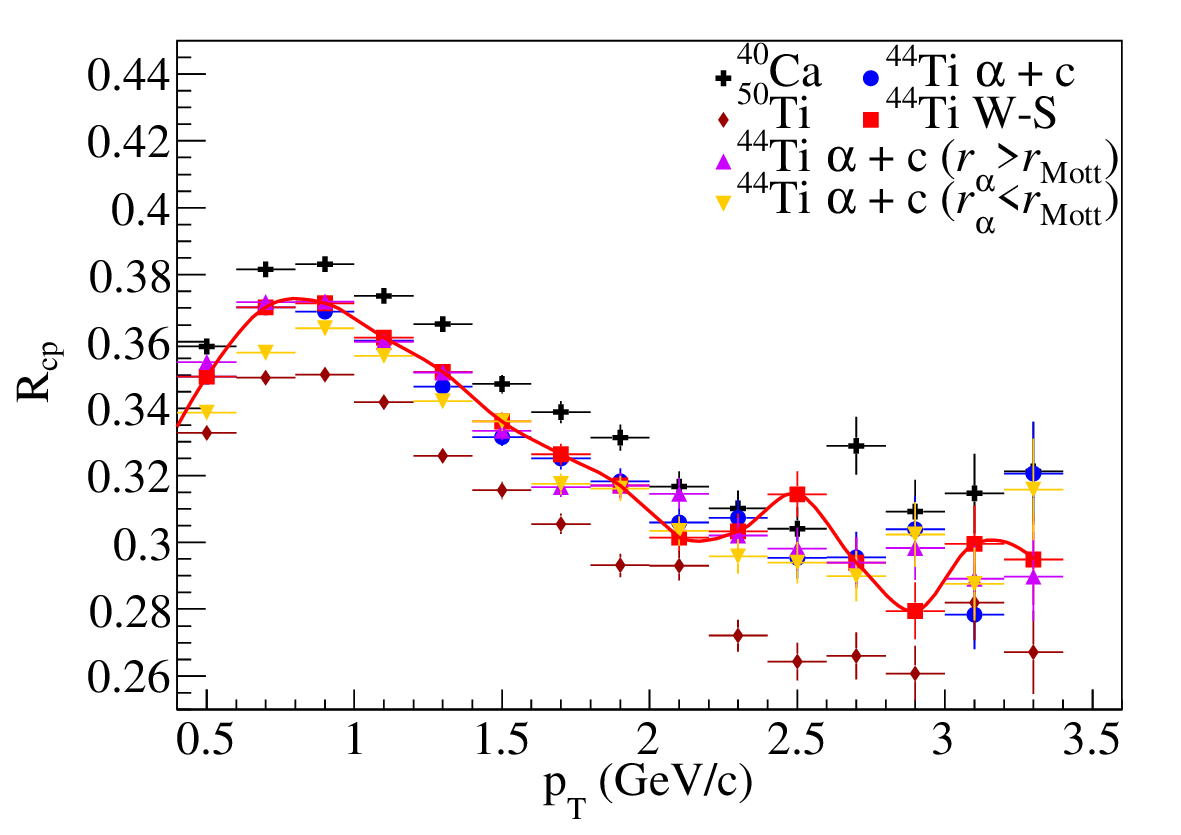}
	\caption{The nuclear modification factor $R_{\mathrm{CP}}$ of the centrality class 0-10\% as a function of $p_{\mathrm{T}}$. The centrality class 80-90\% is the denominator.} 
	\label{fig:graph_rcp}
\end{figure} 
\begin{figure}[tb]
	\includegraphics[width=8.6cm]{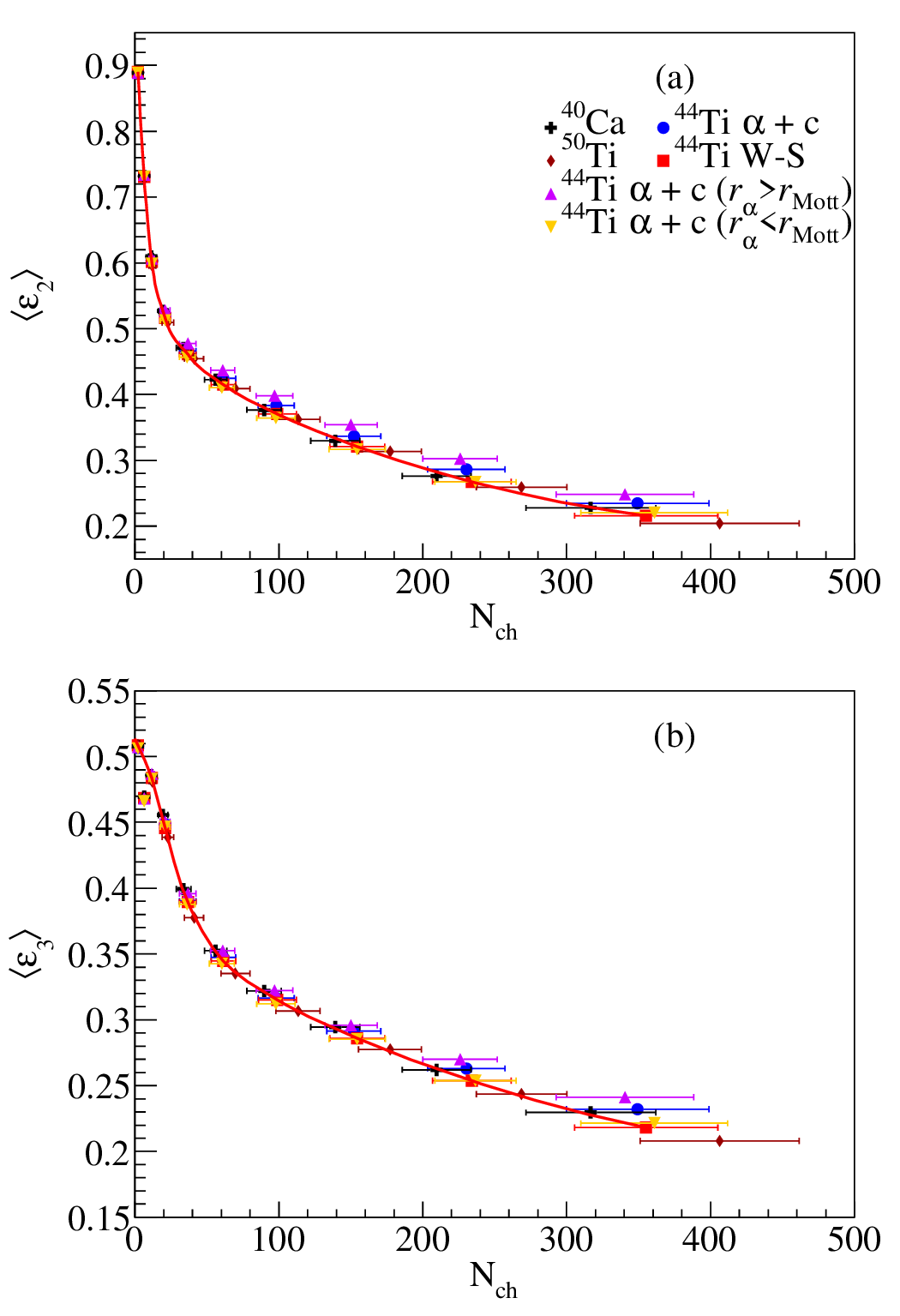}
	\caption{Eccentricity (a) and triangularity (b) of participant partons as a function of $N_{\mathrm{ch}}$.} 
	\label{fig:graph_ir_multi}
\end{figure} 

To further look into this phenomenon, we introduce the nuclear modification factor $R_{\mathrm{CP}}$, which can be defined as \cite{2015392}
\begin{eqnarray}\label{rcp}
	R_{\mathrm{CP}}=\frac{\mathrm{d}^{2}N^{\mathrm{cent}}_{\mathrm{ch}}/(\mathrm{d}p_{\mathrm{T}}\mathrm{d}y)/\left<N^{\mathrm{cent}}_{\mathrm{coll}}\right>}{\mathrm{d}^{2}N^{\mathrm{peri}}_{\mathrm{ch}}/(\mathrm{d}p_{\mathrm{T}}\mathrm{d}y)/\left<N^{\mathrm{peri}}_{\mathrm{coll}}\right>}.
\end{eqnarray}

\begin{figure*}[tb]
	\includegraphics[width=17.8cm]{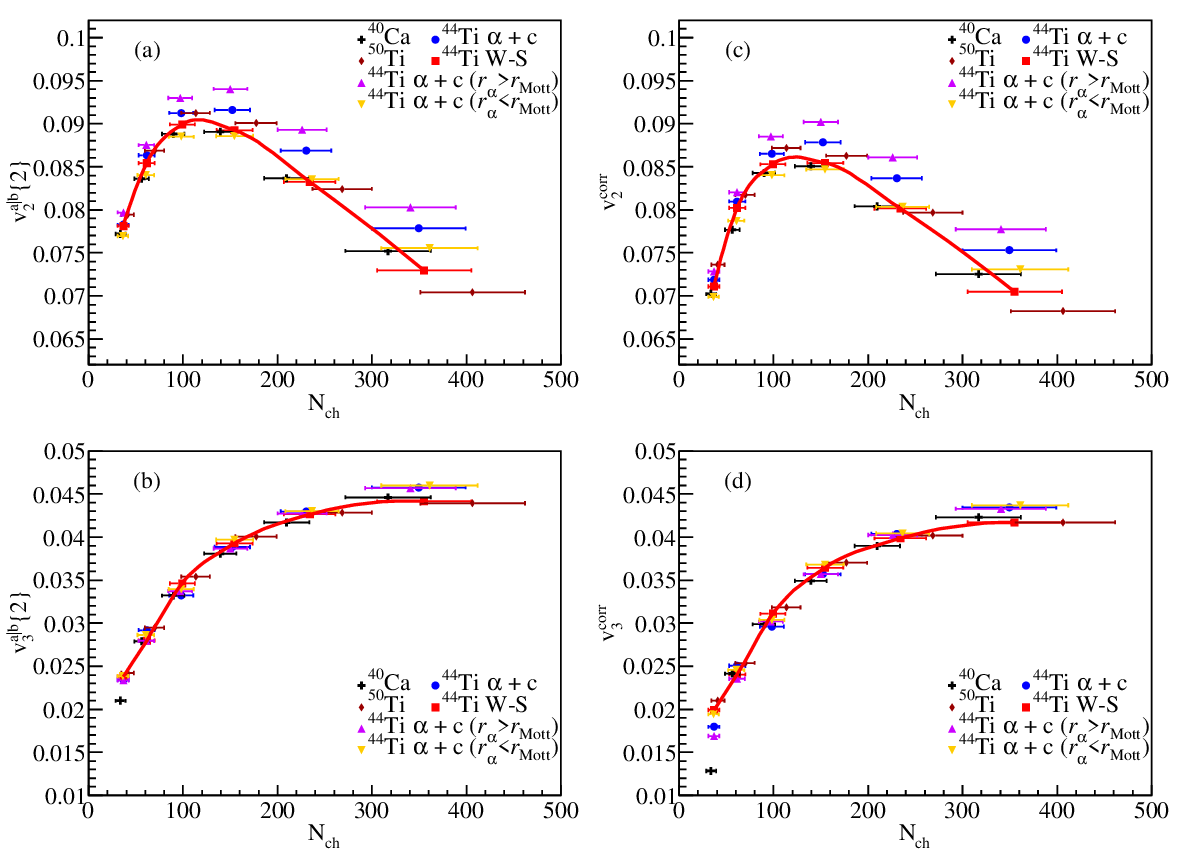}
	\caption{The elliptic flow and triangular flow by the two sub-event cumulant method or by the direct two-particle correlation method as a function of $N_{\mathrm{ch}}$. (a) and (b) are $v_{2}^{a|b}\{2\}$ and $v_{3}^{a|b}\{2\}$ by the two sub-event cumulant method, and (c) and (d) are $v_{2}^{corr}$ and $v_{3}^{corr}$ by the direct two-particle correlation method. } 
	\label{fig:graph_v_multi}
\end{figure*}

\begin{figure}[tb]
	\includegraphics[width=8.6cm]{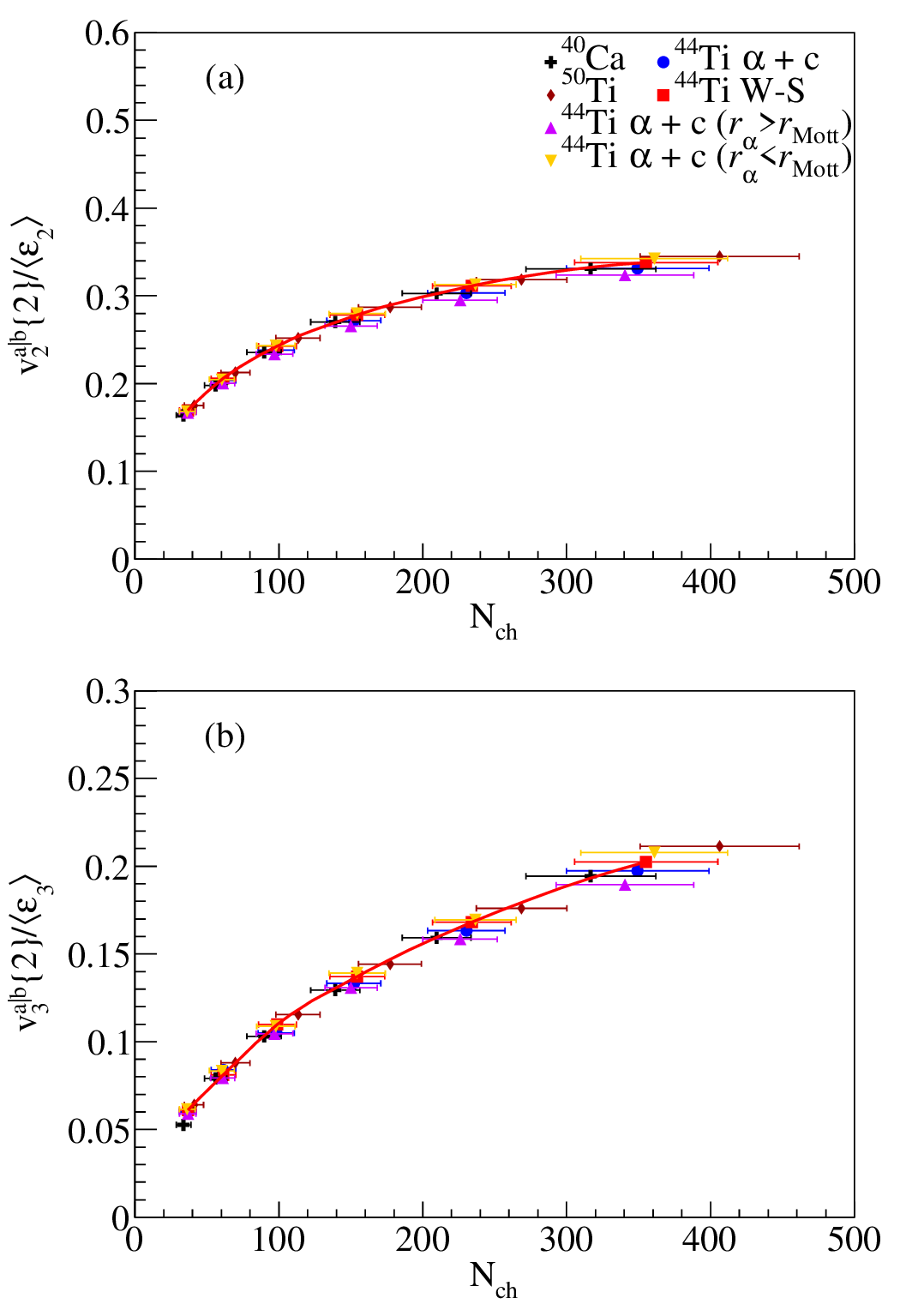}
	\caption{Ratios of $v_{2}^{a|b}\{2\}$ to $\left<\varepsilon_{2}\right>$ (a) and $v_{3}^{a|b}\{2\}$ to $\left<\varepsilon_{3}\right>$ (b) as a function of $N_{\mathrm{ch}}$.} 
	\label{fig:graph_v_ratio_multi}
\end{figure} 

\begin{figure}[tb]
	\includegraphics[width=8.6cm]{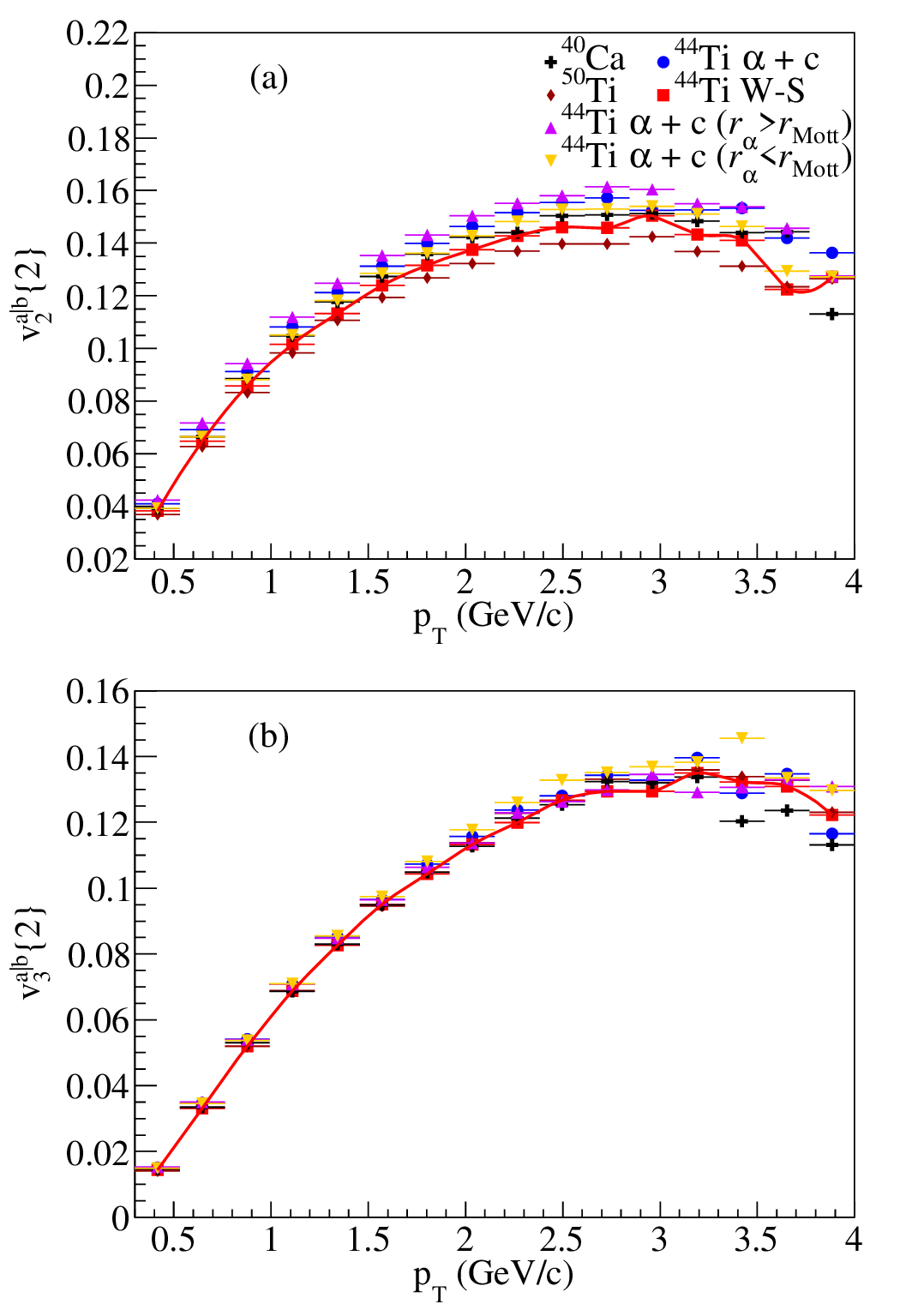}
	\caption{$v_{2}^{a|b}\{2\}$ (a) and $v_{3}^{a|b}\{2\}$ (b) as a function of $p_{\mathrm{T}}$ in the centrality class 0-10\%.} 
	\label{fig:graph_v_0}
\end{figure} 

Here $\left<N_{\mathrm{coll}}\right>$ is the average inelastic binary collision number over events. The superscript ``cent" represents central collisions and ``peri" means peripheral events. $R_{\mathrm{CP}}$ is usually used to estimate the particle production suppression, and is widely considered to be negatively correlated with multiplicity. Figure \ref{fig:graph_rcp} shows the $R_{\mathrm{CP}}$ results of the centrality class 0-10\% with the class 80-90\% as the denominator. The $^{40}$Ca + $^{40}$Ca system shows the highest $R_{\mathrm{CP}}$, while $^{50}$Ti + $^{50}$Ti has the lowest, reflecting typical multiplicity correlation. Among $^{44}$Ti + $^{44}$Ti systems, in the low $p_{\mathrm{T}}$ range the difference caused by nuclear structures is more significant than in Fig. \ref{fig:sp}, which is also negatively correlated with the multiplicity. At other $p_{\mathrm{T}}$, $R_{\mathrm{CP}}$ is almost the same for all $^{44}$Ti + $^{44}$Ti systems. Unfortunately, such difference is still far from distinguishing itself compared with the system size influence.

Another issue about the initialization configuration is the collision zone shape, which can be characterized by $\varepsilon_{n}$ defined in Eq.~(\ref{eccentricity}). Fig. \ref{fig:graph_ir_multi} presents $\left<\varepsilon_{2}\right>$ and $\left<\varepsilon_{3}\right>$ of participant partons in all centrality classes as a function of $N_{\mathrm{ch}}$. $\left<\varepsilon_{2}\right>$ and $\left<\varepsilon_{3}\right>$  decrease with the increasing of $N_{ch}$ in these collision systems, which presents a system size dependence  like that in our previous work~\cite{ZHANG2020135366}. For Woods-Saxon nuclei, their eccentricity follows an order of multiplicity, implying that for similar systems $\varepsilon_{n}$ is mainly affected by the system size. For cluster + core nuclei, at low $N_{\mathrm{ch}}$ the eccentricity is almost the same as for the Woods-Saxon structure, while at high $N_{\mathrm{ch}}$ an eccentricity enhancement appears when $\alpha$ clusters are in $r_{\alpha}>r_{\mathrm{Mott}}$ or have no $r_{\alpha}$ constraint. The enhancement is stronger in $r_{\alpha}>r_{\mathrm{Mott}}$, suggesting that it is $\alpha$ clusters outside the cores that cause this $\varepsilon_{n}$ enhancement. These behaviors of $\varepsilon_{n}$ also match the analysis of particle production above, where the influence of cluster + core structures plays an important role in high multiplicity events but diminishes in peripheral collisions.   Interestingly, the $\alpha$-cluster structure effects on the eccentricity are consistent with our previous investigations~\cite{PhysRevC.95.064904,PhysRevC.102.054907} for $^{12}\mathrm{C}+^{12}\mathrm{C}$ and $^{16}\mathrm{O}+^{16}\mathrm{O}$ collisions where the chain structure enhances $\varepsilon_2$ and reduces $\varepsilon_3$ and the triangle or tetrahedron structure do the opposite. This implies that the nuclear intrinsic structure plays an important role in the geometry shape formation at the collision initial stage.

Figure \ref{fig:graph_v_multi} shows the elliptic flow and the triangular flow of major charged particles gained with two sub-event cumulants or direct two-particle correlation. For the two sub-event cumulant method, the kinetic window is $-2.4<\eta_{a}<0$, $0<\eta_{b}<2.4$ and $0.3<p_{\mathrm{T}}<4.0~\mathrm{GeV/c}$, where $\eta_{a}$ and $\eta_{b}$ are the pseudorapidity of particles in sub-events a and b, respectively. For direct two-particle correlation, the kinetic window of the trigger and associated particles is $0.3<p_{\mathrm{T}}<4.0~\mathrm{GeV/c}$ and $-2.4<\eta<2.4$, so that the measurement involves the same particles as in the two sub-event cumulant method. A pseudorapidity gap $\Delta\eta>2.0$ is required for effective correlation to avoid non-flow effects. The result of peripheral events in centrality $60-100\%$ is omitted since jet-like correlation is the main source of $v_{2}$, and $v_{3}$ here. In Fig. \ref{fig:graph_v_multi} (a) $v_{2}^{a|b}\{2\}$ shows obvious enhancement for cluster + core structures except in $r_{\alpha}<r_{\mathrm{Mott}}$, while in Fig. \ref{fig:graph_v_multi} (b) $v_{3}^{a|b}\{2\}$ of all systems follows similar multiplicity order except in the highest multiplicity class. In Fig. \ref{fig:graph_v_multi} (c) and (d), the results of $v_{2}^{corr}$ and $v_{3}^{corr}$ show great similarity with $v_{2}^{a|b}\{2\}$ and $v_{3}^{a|b}\{2\}$, showing that the difference between $v_{2}$ and $v_{3}$ is independent of the exact tool to extract the flow. 

To carry out further investigation, we try a new perspective of the flow-to-eccentricity ratio. Hydrodynamic calculations \cite{NSTSong2017,PhysRevC.84.024911} point out that there is a proportional relationship between $\varepsilon_{n}$ and $v_{n}$ for $n=2,3$, so their ratio might be able to reflect the efficiency to transform initial geometry asymmetry to final momentum space asymmetry \cite{PhysRevC.102.054907}. The ratios of $v_{n}^{a|b}\{2\}$ to $\left<\varepsilon_{n}\right>$ are shown in Fig. \ref{fig:graph_v_ratio_multi}. Although the behaviors of $v_{2}^{a|b}\{2\}$ and $v_{3}^{a|b}\{2\}$ are quite different, their ratios to corresponding eccentricity do share the same pattern. Like the case of $\left<\varepsilon_{n}\right>$, $v_{n}^{a|b}\{2\}/\left<\varepsilon_{n}\right>$ of Woods-Saxon systems follows the order of multiplicity, while for core + cluster nuclei, $v_{n}^{a|b}\{2\}/\left<\varepsilon_{n}\right>$ shows influence of nuclear structures at high $N_{\mathrm{ch}}$ except in $r_{\alpha}<r_{\mathrm{Mott}}$. This time binary structures lead to a slight depression, indicating that the cluster effects may be weakened by particle interaction during collisions to some extent. Combining $\left<\varepsilon_{n}\right>$, $v_{n}^{a|b}\{2\}$ and $v_{n}^{a|b}\{2\}/\left<\varepsilon_{n}\right>$, we can say that the different behaviors between $v_{2}$ and $v_{3}$ are likely to be the natural result of system evolution. 

To investigate the kinetic window dependence of the influence of $\alpha$-cluster nuclear structure, the $p_{\mathrm{T}}$-differential $v_{n}^{a|b}\{2\}$ of the centrality class 0-10\% is calculated and shown in Figure \ref{fig:graph_v_0}. $v_{2}^{a|b}\{2\}$ and $v_{3}^{a|b}\{2\}$ increase with the increasing of $p_T$ and reach the maximum around $p_T\sim$ 2.5 GeV/$c$, and then present a decreasing tend. The core + cluster enhancement is still obvious this time for $v_{2}^{a|b}\{2\}$, while $v_{3}^{a|b}\{2\}$ remains similar for all six systems. So the $\alpha$-cluster nuclear structure effect on collective flow dose not depend on the selected $p_T$ window and $p_{\mathrm{T}}$-differential collective flow may also be a good probe for nuclear structures.

From the above analysis of the differential flow in the most-central collisions and the $N_{ch}$ dependence of flow, it is shown that elliptic flow coefficients in the natural nucleus collision systems, $^{50}$Ti + $^{50}$Ti, $^{44}$Ti(W-S) + $^{44}$Ti(W-S), $^{40}$Ca + $^{40}$Ca, follow the order of $v_2(^{40}\mathrm{Ca})$ $>$ $v_2(^{44}\mathrm{Ti(W-S)})$ $>$ $v_2(^{50}\mathrm{Ti})$ which is consistent with our previous study of the system size dependence of collective flow~\cite{ZHANG2020135366}. And it is obvious that $^{44}$Ti + $^{44}$Ti configured with $\alpha$-clusters is significantly biased against the order of $v_2$ in the natural nucleus collision systems, up to 10\%, no matter whether the Mott density is taken into account for $^{44}$Ti. In the experiment, a system scan of relativistic heavy-ion collisions near $^{44}$Ti could  be considered as a probe to explore the signature of the $\alpha$-cluster structure in $^{44}$Ti.

\section{Conclusion}\label{Conclusion}

In this work, we test the influence of $\alpha$ cluster + core structures on particle production and collectivity in $^{44}$Ti + $^{44}$Ti collisions at $\sqrt{s_{NN}}=5.02$ TeV with the AMPT model where the initial geometry properties is configured by using  the traditional treatment method of the nuclear structure. Due to $\alpha$ dissolution, there might be the so-called Mott density restricting the positions of $\alpha$ clusters. So we try three modes of binary structures, which are $r_{\alpha}<r_{\mathrm{Mott}}$, $r_{\alpha}>r_{\mathrm{Mott}}$ and free $r_{\alpha}$. For real nuclei, only the last two are possible. The absence of $\alpha$ dissolution greatly increases central nucleon density and leads to the breaking of nuclear matter saturation, implying the necessity of including Mott density in future $\alpha$ + core structure research. Clusters outside cores also raise the peripheral nucleon density but to a much smaller degree.

In particle production, the structure influence mainly contributes to high multiplicity events. Systems with clusters inside cores show a higher fraction of high $N_{\mathrm{ch}}$ events, and clusters outside cores lead to the contrary. The mixed mode is similar to the Woods-Saxon situation, with a much weaker enhancement at high $N_{\mathrm{ch}}$. The alteration is less than one order of magnitude, so it is hard to detect binary structures in transverse momentum spectra. We also calculate the nuclear modification factor $R_{\mathrm{CP}}$ in the centrality class 0-10\% and found that  the signal is stronger, but still not enough to decide the structure. The initial geometry is also affected by cluster + core structures. Both $\varepsilon_{2}$ and $\varepsilon_{3}$ are increased by clusters outside cores, while clusters inside seem to have no effect. The difference in $v_{2}$ is strengthened in comparison  with $\varepsilon_{2}$, so there may be possibility to distinguish the binary structure with a cross-check of elliptic flow in similar systems. However, in $v_{3}$ the difference almost vanishes due to system evolution, leaving traces only in central events. 

Although it is still under debate whether $\alpha$ cluster structures can generate strong signatures in ultra-relativistic heavy-ion collisions, there have been many predictions about the all-cluster structures. Our result provides a new insight of $\alpha$ + core structures into this issue, showing that elliptic flow might be a good probe of binary structures, while triangular flow and particle production are less sensitive to these structures. And a system scan of relativistic heavy-ion collisions is proposed to be performed in the experiment to investigate $\alpha$ + core structures through the system dependence of elliptic flow.

\begin{acknowledgments}
This work was supported in part by National Key R\&D Program of China under Grant No. 2018YFE0104600 and 2016YFE0100900, the National Natural Science Foundation of China under contract Nos. 12275054, 12147101, 11925502, 11890710, 11890714, 12061141008 and 11875066, the Strategic Priority Research Program of CAS under Grant No. XDB34000000, the Guangdong Major Project of Basic and Applied Basic Research No. 2020B0301030008, Shanghai Special Project for Basic Research No. 22TQ006 and the STCSM under Grant No. 23590780100.	
\end{acknowledgments}

\bibliography{ref}
\end{document}